\documentclass[10pt,pra,aps,twocolumn,superscriptaddress,showpacs]{revtex4-1}
\usepackage{amsmath,amsfonts,amssymb}
\usepackage{braket} 
\usepackage{ulem} 
\usepackage{framed} 
\usepackage[pdftex]{graphicx}
\usepackage{natbib}
\usepackage{bbm}    
\usepackage{color}
\usepackage{xcolor}
\usepackage{multirow} 

\newcommand{\bs}[1]{\boldsymbol{#1}}

\begin{document}

\title{Transmission through a potential barrier in Luttinger liquids with a topological spin gap }

\author{Nikolaos Kainaris}
\affiliation{Institut f\"ur Nanotechnologie, Karlsruhe Institute of Technology, 76021 Karlsruhe, Germany}
\affiliation{\mbox{Institut f\"ur Theorie der Kondensierten Materie, Karlsruher Institut f\"ur
  Technologie, 76128 Karlsruhe, Germany}}


\author{Sam T. Carr} 
\affiliation{School of Physical Sciences, University of Kent, Canterbury CT2  7NH, United Kingdom}

\author{Alexander D. Mirlin}
\affiliation{Institut f\"ur Nanotechnologie, Karlsruhe Institute of Technology, 76021 Karlsruhe, Germany}
\affiliation{\mbox{Institut f\"ur Theorie der Kondensierten Materie, Karlsruher Institut f\"ur
  Technologie, 76128 Karlsruhe, Germany}}
\affiliation{Petersburg Nuclear Physics Institute,
 188350 St. Petersburg, Russia}


\begin{abstract}
\noindent
We study theoretically the transport of the one-dimensional single-channel interacting electron gas through a strong potential barrier in the parameter regime where the spin sector of the low-energy Luttinger liquid theory is gapped by interaction. This phase is of particular interest since it exhibits non-trivial interaction-induced topological properties. Using bosonization and an expansion in the tunneling strength, we calculate the conductance through the barrier as a function of the temperature as well as the local density of states (LDOS) at the barrier. Our main result concerns the mechanism of bound-state mediated tunneling. The characteristic feature of the topological phase is the emergence of protected zero-energy bound states with fractional spin located at the impurity position. By flipping the fractional spin the edge states can absorb or emit spinons and thus enable single electron tunneling across the impurity even though the bulk spectrum for these excitations is gapped. This results in a finite LDOS below the bulk gap and in a non-monotonic behavior of the conductance. The system represents an important physical example of an interacting symmetry-protected topological phase---which combines features of a topological spin insulator and a topological charge metal---in which the topology can be probed by measuring transport properties.

\end{abstract}

\maketitle

\section{Introduction}
\label{sec:introduction}

Following the prediction and experimental discovery of topological insulator materials over the last decade~\cite{Kane_Mele_2005,Bernevig_2006,Moore_2007,Fu_2007,Roy_2009,Koenig_2007,Hsieh_2008}, much of the experimental and theoretical effort in recent years has turned towards investigating related topological phenomena in strongly correlated materials. Major progress in understanding these phases has been achieved for systems in one spatial dimension, where a formal mathematical classification of all symmetry protected phases has been developed~\cite{Pollmann_2010, Fidkowski_2011, Turner_2011,Schuch_2011, Chen_2011}. More recently, a generalization of these methods to systems with dimensionality $d>1$ was proposed ~\cite{Chen_2012, Chen_2013, Gu_2014}.


While the complete classification of one-dimensional (1D) symmetry-protected topological phases  has constituted an important breakthrough, it is not sufficient by itself to determine physical properties of such systems. In particular, predictions for transport properties of strongly-correlated topological materials are highly desirable as they often
offer the most straightforward way to experimentally probe for systems with nontrivial topology.

An important model system for studying transport properties in a strongly correlated symmetry-protected topological phase is the one-dimensional (1D) electron gas with time reversal symmetry, electron-electron interaction, and spin-anisotropy~\cite{Keselman_2015, Kainaris_Carr_2015, Haim_2016, Montorsi_2017, Kainaris_2017}.
As is well known, the spin and charge degrees of freedom of the 1D electron gas decouple in the low-energy Luttinger liquid theory which describes the behavior of gapless collective bosonic excitations~\cite{Giamarchi_book,Gogolin_book}. In a certain parameter range, the modes in the spin sector get dynamically gapped out due to electron-electron backscattering processes which grow under the renormalization flow. On the other hand, the charge sector  remains gapless. Without any spin anisotropy, this gapped phase is characterized by quasi-long-range charge density wave (CDW) correlations and was shown to be  topologically trivial~\citep{Kainaris_Carr_2015}. If, however, the spin anisotropy is large enough, the system flows to a different phase which shows nontrivial topological features. This topological phase exhibits quasi-long range spin density wave (SDW) correlations in the bulk and zero-energy boundary bound states (BBS) which carry fractional spin~\citep{Keselman_2015,Kainaris_Carr_2015}. The peculiarity of both the SDW and CDW phase is that the excitations in the spin sector are gapped while the charge sector remains gapless. As such these phase are inherently distinct from non-interacting topological insulators, where both sectors are gapped. On the other hand, they are also distinct from each other since only the SDW phase shows topologically nontrivial features.

\begin{figure}
  \centering
  \includegraphics[width=.45\textwidth]{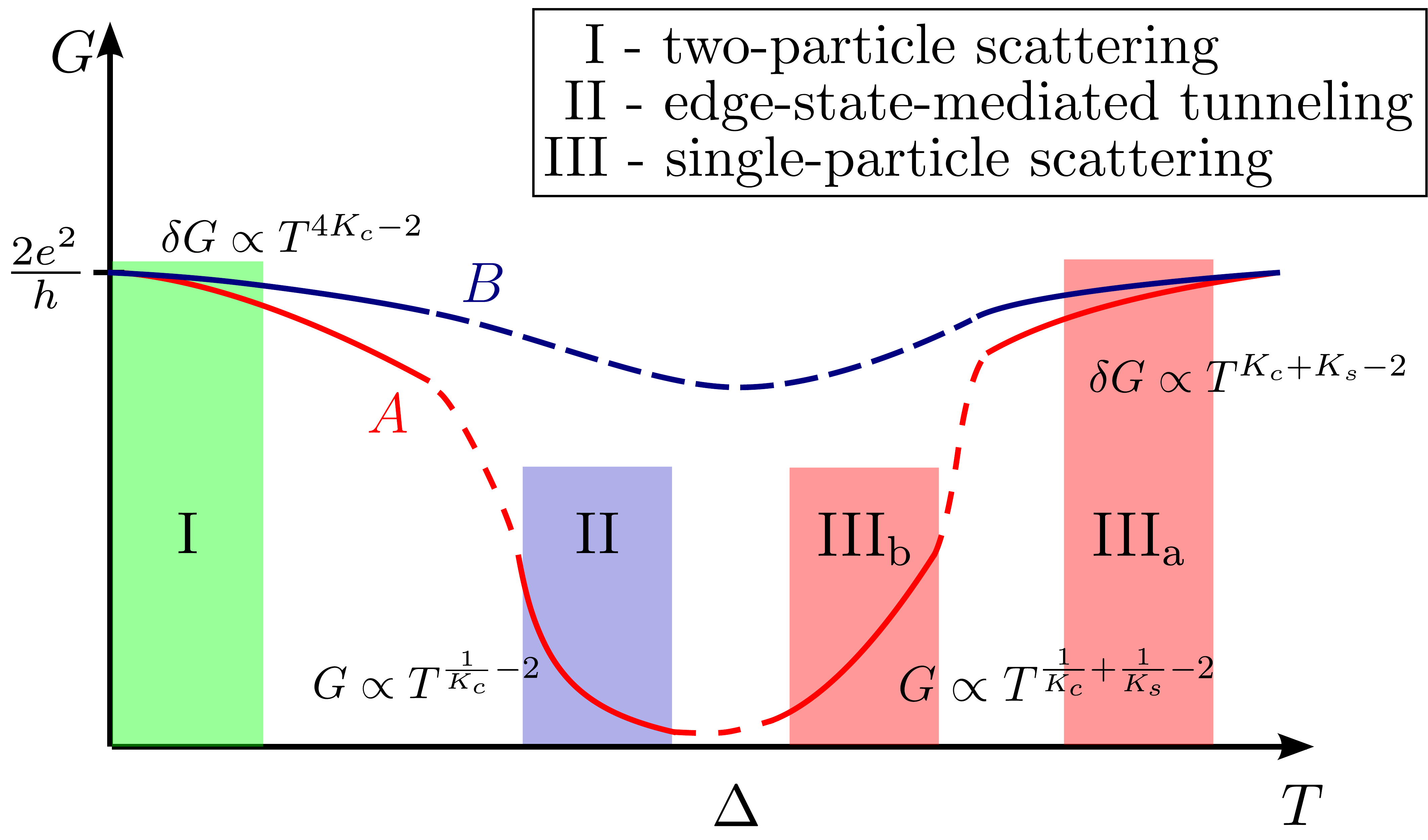}
  \caption{Sketch of the temperature dependence of the conductance of a topological spin-anisotropic 1D electron system in the presence of an impurity. Regime III corresponds to the Luttinger liquid phase, with the subscripts $a$ and $b$ denoting a weak and strong impurity potential, respectively.
 Regimes I and II correspond to the topologically gapped SDW phase with a weak and strong potential barrier, respectively. 
   The lines denoted by A and B show the behavior of the conductance for two sets of parameters. In the case B, the impurity remains weak under the RG flow and the transport behavior crosses over from region $\text{III}_\text{a}$ to I directly as temperature is lowered. On the other hand, in the case A,  the impurity potential becomes strong under the RG flow, which leads to multiple crossovers $\text{III}_\text{a} \to \text{III}_\text{b} \to \text{II} \to \text{I} $ between regimes with different transport characteristics. The transport mechanisms in different regimes are discussed in detail in the main text. }
  \label{Fig:conductance_schematic}
\end{figure}

\begin{figure}
  \centering
  \includegraphics[width=.35\textwidth]{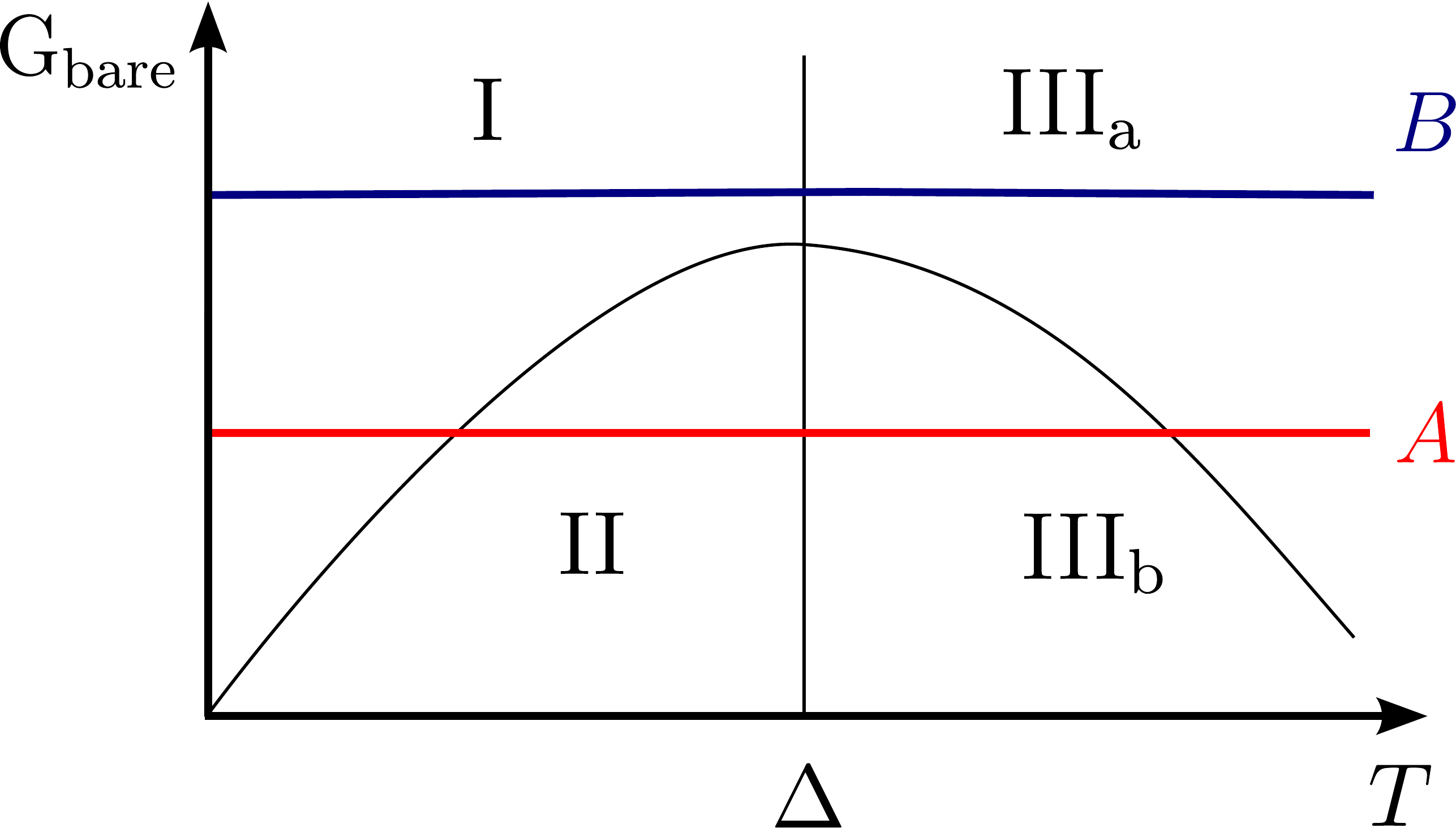}
  \caption{Diagram of transport regimes of a topological spin-anisotropic 1D electron system in the presence of an impurity in the parameter plane spanned by the bare conductance $G_{\text{bare}}$ and the temperature $T$. The regimes I-III are defined in the main text and in the caption of Fig.~\ref{Fig:conductance_schematic}. The red and blue lines (marked by A and B, respectively) correspond to an impurity that is initially weak or strong, respectively; the corresponding temperature dependences of the conductance are sketched in Fig.~\ref{Fig:conductance_schematic}. }
  \label{Fig:conductance_phase_diagram}
\end{figure}

The subject of this paper is the transport through an impurity in the topological phase described above. 
It turns out that, in addition to the nontrivial boundary spectrum, the topological SDW phase also exhibits novel transport properties distinct from both the conventional Luttinger liquid and trivial CDW phase. In particular the bulk transport of the system remains ballistic in the low-temperature limit even in the presence of impurities as long as the time reversal symmetry is preserved and interactions are not too strong. More precisely, a single impurity acts as an irrelevant perturbation as long as $K_c>1/2$, where $K_c$ denotes the Luttinger liquid parameter in the charge sector.

The schematic behavior of the bulk conductance of the SDW phase in presence of a single nonmagnetic scatterer as a function of the temperature is shown in Fig.~\ref{Fig:conductance_schematic}. This figure combines the results of the present work with previously known results.
An overview of the different transport regimes of the system is depicted in Fig.~\ref{Fig:conductance_phase_diagram}.

In region III, which is the regime of temperatures much higher than the spin gap $\Delta$, the system effectively behaves as a Luttinger liquid in the presence of a single impurity. The transport properties in this regime are well known~\cite{Kane_Fisher_1992, Furusaki_Nagaosa_1993}: for repulsive interactions the impurity represents a relevant perturbation which causes the conductance to decrease as a power law as temperature is lowered. The exponent of this power law differs depending on whether the impurity potential is weak (region $\text{III}_{\text{a}}$) or strong (region $\text{III}_{\text{b}}$). The region I, which describes a weak impurity at temperatures $T \ll \Delta$, has been analyzed by two of us in Ref.~\cite{Kainaris_Carr_2015}. It was found that the conductance at lowest temperatures behaves ballistic with small power-law corrections at finite temperature. Physically, these corrections stem from scattering of singlet electron pairs off the impurity: due to the excitation gap for spin-$1/2$ particles, the lowest energetically allowed excitations are electron spin-singlet pairs.

Thus, an impurity may become strong under the renormalization group (RG) in the range of relatively high temperatures $T\gtrsim \Delta$, regime $\text{III}_{\text{b}}$. Contrary to this, a weak impurity becomes weaker at low temperatures, $T \ll \Delta$. This poses a question of the properties of the topological phase ($T \ll \Delta$) with a strong impurity. The analysis of this regime---which is denoted by II in Figs.~\ref{Fig:conductance_schematic} and \ref{Fig:conductance_phase_diagram}---constitutes the main subject of the present work.

The intermediate regime II that we explore here represents a vicinity of a strong-coupling fixed point, where both the impurity strength and the bulk gap have flown to strong coupling under the RG. Using the weakness of the electron tunneling across the barrier in this regime, we will determine the temperature dependence of the conductance as well as the tunneling density of states near the edge.

Our central findings concern the transport mechanism in the regime II and the associated physical observables. On the one hand, we find that although spin-$1/2$ excitations are gapped in the bulk, single electron tunneling can take place via flipping the fractional spin of the boundary bound states.
On the other hand, the bound states are energetically split due to the finite tunneling amplitude, with the energy splitting growing proportionally to the tunneling. As tunneling increases with lowering temperature, there exists a critical scale where the energy splitting becomes of the order of the bulk gap. At this energy scale, the single particle tunneling becomes frozen out and a crossover to pair-tunneling-mediated transport (regime I) occurs. The overall temperature dependence of the conductance is strongly non-monotonous, see Fig.~\ref{Fig:conductance_schematic}. The underlying physics of the problem is also visible in the behavior of the local density of states (LDOS), which has a finite subgap contribution due to the edge states that is gradually shifted in energy towards the bulk density of states  as temperature is lowered. The finite subgap LDOS and the nonmonotonic behavior of the conductance may thus serve as experimental probes of a nontrivial topology of the system.

The paper is organized as follows. In Sec.~\ref{Sec:The model} we introduce the model we are going to study. Next, we calculate the tunneling conductance across the impurity potential barrier to leading order in the tunneling amplitude in Sec.~\ref{Sec:Tunneling current}. To gain more physical insight into the transport results, we show that the model exhibits boundary bound states and calculate their energy as a function of the tunneling strength across the impurity in Sec.~\ref{sec:Boundary bound state}.  Finally, we study the local density of states in Sec.~\ref{Sec:Local density of states} and summarize our findings in Sec.~\ref{Sec:Conclusion}. For completeness, we present in Appendix~\ref{appendix-RG} the renormalization-group analysis and the phase diagram of the model under consideration in the whole range of interaction couplings.

\section{The model}
\label{Sec:The model}

Let us introduce the model we are going to study. We consider an interacting 1D electron gas in the presence of short-range electron-electron interaction and a single impurity located at the origin. The effective low-energy Hamiltonian of this model can be expressed in bosonized language as $H = H_{c} + H_s + H_{\text{imp}}$, with
\begin{align}
   H_c =&  \frac{v_c}{4 \pi K_c}  \int_{-\infty}^{\infty} \! \mathrm{d} x \, \Big[( \partial_x \Phi_{c } )^2 + K_{c}^2(\partial_x \theta_c)^2\Big] \label{Hc}~,\\
\begin{split} 
   H_{s} =& \frac{v_s}{4 \pi K_s}  \int_{-\infty}^{\infty} \! \mathrm{d} x \, \Big[( \partial_x \Phi_{s } )^2 + K_{s}^2(\partial_x \theta_{s})^2\Big] \\
                  &  +  \frac{g_{\text{sG}}}{(2 \pi a)^2} \int_{-\infty}^{\infty} \! \mathrm{d} x \, \cos 2 \Phi_{s}~, \label{Hs}\end{split}   \\
   H_{\text{imp}}  =&  - \frac{g_b}{\pi a} \int_{-\infty}^{\infty} \! \mathrm{d} x \, \cos \Phi_{s}  \cos  \Phi_c \, \delta(x)~.  \label{Himp}
\end{align}
The bosonic operators $\Phi_{c,s}$ and $\theta_{c,s}$ in the charge and spin sectors are related to fermionic operators, describing modes linearized around the Fermi momentum $k_F$, as
\begin{align}
   \psi_{\eta,\sigma} = \frac{\kappa_{\sigma}}{\sqrt{2 \pi a}} e^{i \frac{\eta}{4} \left[ \Phi_c+ \sigma \Phi_s - \eta \theta_c - \eta \sigma \theta_s \right]} \, .
\end{align} 
Here $\sigma= \uparrow ,\downarrow = +,-$ denotes the electron spin, $\eta = +,- $ the electron chirality, $\kappa_{\sigma}$ are Klein factors which ensure the correct fermionic anticommutation relations, and $a$ is the short-distance cutoff of the theory.

The Hamiltonian in the charge sector, Eq.~(\ref{Hc}), describes collective gapless excitations with charge $\pm e$ and velocity $v_c$. The Luttinger parameter $K_c$ is a measure of the strength of electron-electron interactions. Note that we consider electrons at incommensurate filling, so no Umklapp terms are present in the model. 

On the other hand, spin excitations in the bulk are described by the sine-Gordon model, Eq.~(\ref{Hs}). Here, the cosine term originates from electronic backscattering processes and can lead to the formation of a gap. More precisely, in the regime $1/2<K_s<1$ the excitations in the spin sector are gapped solitons and antisolitons that carry spin $1/2$ and $-1/2$, respectively. As was pointed out in Ref.~\citep{Kainaris_Carr_2015}, the system can also flow to this gapped fixed point dynamically for $K_s>1$ if Ising spin anisotropy is present. Throughout this work we restrict ourselves to the parameter regime $K_s \geq 1/2$. For $K_s<1/2$ propagating breather (soliton-antisoliton) bound states would exist which are not considered here, although we would not expect this to qualitatively change the properties of the state that we discuss.

For completeness, we derive the phase diagram of the model Eqs.~(\ref{Hc})-(\ref{Himp}) in Appendix \ref{appendix-RG}.  For the purposes of the rest of the text however, it is sufficient to say that we are in a phase where the bulk cosine term in Eq.~\eqref{Hs} is relevant and the system develops a spin gap, while the charge sector of the theory remains gapless.  Such a phase is often termed as a Luther-Emery liquid \cite{Luther_Emery_1974}.

Thermodynamically, the sign of the coupling constant $g_{\text{sG}}$ is not important, and there is a duality $g_\text{sG}\rightarrow -g_\text{sG}$.  It should be stressed however that the topological nature of the gapped phase depends crucially on the sign of this coupling constant. In fact, one can define a topological index $\mathcal{Q} = \text{sgn}(g_{\text{sG}})$ which takes the value $\mathcal{Q} = +1$ in the topological and $ \mathcal{Q} = -1$ in the topologically trivial phase. Throughout this paper we will assume $g_{\text{sG}} > 0$ since we are interested in studying the topological phase.

Lastly, the term (\ref{Himp}) in the Hamiltonian describes a time-reversal-symmetric impurity potential with a strength $g_b$. We assume $g_b>0$  but actually the physical results do not depend on the sign of $g_b$. Let us note that the term (\ref{Himp}) mixes the charge and spin sectors. 

While we will discuss the model defined by Eqs.~(\ref{Hc})--(\ref{Himp}) in the context of spinful electrons~\cite{Kainaris_Carr_2015}, we note that other physical systems are also described by the same low-energy Hamiltonian. Examples include cold-atom systems~\cite{Iemini_2017}, coupled superconducting wires~\cite{Fidkowski_2011a,Sau_2011,Keselman_2015}, coupled edges of quantum spin Hall insulators~\cite{Santos_2016, Kainaris_2017}, ladder models~\cite{Cheng_2011,Carr_2011}, and Kondo chains~\cite{Tsvelik_2015,Schimmel_2016}.  We also refer to these earlier works for the relationships between the parameters in the low energy effective theory  Eqs.~(\ref{Hc})-(\ref{Himp}) and any given microscopic system.

\section{Tunneling current across a large potential barrier}
\label{Sec:Tunneling current}

In this Section we will discuss the transport properties of the model introduced in Eqs.~(\ref{Hc})--(\ref{Himp}) in the regime II of Fig.~\ref{Fig:conductance_schematic}. To be more precise, we want to analyze the conductance in the regime where both the impurity potential, $g_b/ v_s \gg 1$, and the bulk interaction potential, $g_{\text{sG}}/ v_s \gg 1 $, are strong. 

First, we derive an effective model in the regime $g_{\text{sG}}/ v_s \gg 1 $.
If correlations in the bulk are strong, the spin field will establish a mean field $\Phi_s^{\text{SDW}} \; = \;  \pi/2$ in order to minimize the potential energy of the bulk term~(\ref{Hs}). Quantum-mechanical fluctuations around this ground state can be described semiclassically by writing $\Phi_s(x,\tau) = \Phi_s^{\text{SDW}} + \delta \Phi_s(x,\tau)$ and expanding the action of the model to quadratic order in fluctuations $\delta \Phi_s$. This yields the following action in energy-momentum space:
\begin{eqnarray}
   &&      S_{\text{LL}} [\Phi_c,\delta \Phi_s] =  \frac{1}{4 \pi v_c K_c} \int_{\omega,q} |\Phi_c(q,\omega)|^2 (\omega^2 +v_c^2 q^2) \nonumber \\
     && \qquad       +  \frac{1}{4 \pi v_s K_s} \int_{\omega,q} |\delta \Phi_s(q,\omega)|^2 (\omega^2 +v_s^2 q^2 + \Delta^2)~,
\label{actionmassivefluct}
      \end{eqnarray}
where $\Delta = \left(8 \pi v_s^2 K_s g_{\text{sG}}\right)^{1/2} /a$ denotes the excitation gap of the spin fluctuations. In terms of the fluctuations, the impurity potential takes the form
      \begin{align}
         S_{\text{imp}}[\Phi_c,\delta \Phi_s] \; = \; & \frac{g_b}{\pi a}\int \!\mathrm{d} \tau \, \cos \Phi_c(0,\tau) \sin \delta \Phi_s(0,\tau) \, . \label{Simp}
      \end{align}   
Next we integrate out all fields except those at the origin to obtain an effective local action:
\begin{align}
   & e^{-S_{\text{eff}}[q_c,q_s]} = \int \! \mathrm{D} \Phi_c \int \! \mathrm{D} \Phi_s \, 
   \delta[q_c(\tau) - \Phi_c(0,\tau) ]  \nonumber \\
&\times    \delta[q_s(\tau)  - \delta \Phi_s(0,\tau) ] e^{-S_{\text{LL}}[\Phi_c,\delta \Phi_s]-S_{\text{imp}}[\Phi_c,\delta \Phi_s]}~. 
\end{align}
Performing the Gaussian functional integration, we arrive at the result
\begin{align}
\begin{split} 
   S_{\text{eff}} =& \sum_{\mu=c,s} \int_{\omega} \mathcal{K}_{\mu}(\omega) |q_{\mu}(\omega)|^2   
                   + S_{\text{imp}}[q_c,q_s]~,  
                   \label{effectiveaction}
\end{split}
\end{align}
with the kernel functions
\begin{align}
   \mathcal{K}_c(\omega) =& \frac{1}{2 \pi K_c}|\omega| ~, \\
   \mathcal{K}_s(\omega) =& \frac{1}{2 \pi K_s} \sqrt{\omega^2+\Delta^2}.
\end{align}
As was pointed out by Furusaki and Nagaosa~\cite{Furusaki_Nagaosa_1993}, this type of action is equivalent to that of a quantum Brownian particle moving in a periodic cosine potential and coupled to a dissipative environment. However, unlike in the gapless Luttinger liquid, only the low-lying charge excitations cause the damping in our model (as long as we are interested only in energies below the gap). The reason for this is that spin excitations are gapped and thus can not contribute to the damping. 

To avoid UV divergencies it is necessary to introduce a high-frequency cutoff  which correspond to a finite mass of the Brownian particle. 
Since we are interested in the physics on energy scales below the gap $\Delta$, we choose the cutoff to be of the order of $\Delta$. 

In the limit of a large impurity potential, the electron transport can be viewed as a tunneling from a minimum of the potential (\ref{Simp}) to an adjacent minimum. The corresponding tunneling amplitude $\gamma$ (at the new ultraviolet cutoff scale $\Delta$) provides a natural expansion parameter. The relationship between the tunneling strength and the barrier strength is non-universal \cite{Aristov_2009}, so we will consider $\gamma$ to be a phenomenological parameter. In the following, we will calculate the conductance perturbatively in leading order in the tunnneling amplitude $\gamma$ by using Fermi's golden rule. The analysis generalizes the discussion of Ref.~\cite{Furusaki_Nagaosa_1993} to the case of a gapped spin sector.

\begin{widetext}
\noindent
It is useful to rewrite the partititon function by introducing a set of quadratic oscillator degrees of freedom $\lbrace x_{1j} \rbrace$ and $\lbrace x_{2k} \rbrace$, 
\begin{align}
   Z &= \int \! \mathrm{D} [q_c,q_s] \, e^{-S_{\text{eff}}[q_c,q_s]} 
     = \prod_{jk} \int \! \mathrm{D} [q_c, q_s,x_{1j},x_{2k}] \, e^{-\int\! \mathrm{d} \tau \, \mathcal{L}( \left\lbrace x_{1j} \right\rbrace, \left\lbrace x_{2k} \right\rbrace, q_c,q_s)}~, \label{Z}
\end{align}
with
\begin{align}
    \begin{split} 
    \mathcal{L} =& \sum_j \Big[ \frac{m_{1j}}{2} \dot{x}_{1j}^2 +\frac{m_{1j}}{2} \omega_{1j}^2 x_{1j}^2 + g_{1j} x_{1j} q_c + \frac{g_{1j}^2}{2 m_{1j} \omega_{1j}^2} q_c^2\Big] \\
         &+ \sum_k \Big[ \frac{m_{2k}}{2} \dot{x}_{2k}^2 +\frac{m_{2k}}{2} \omega_{2k}^2 x_{2k}^2 + g_{2k} x_{2k} q_s + \frac{g_{2k}^2}{2 m_{2k} \omega_{2k}^2} q_s^2\Big] + \frac{g_{b}}{\pi a} ( \cos q_c \sin q_s -1 )~. 
    \end{split} \label{lagrangian}
\end{align}
The introduced oscillators are characterized by the spectral functions
\begin{align}
   J_c(\Omega) \; =& \; \frac{\pi}{2} \sum_{j} \frac{g_{1j}^2}{m_{1j} \omega_{1j}} \delta(\Omega-\omega_{1j})~, \\
   J_s(\Omega) \; =& \; \frac{\pi}{2} \sum_{k} \frac{g_{2k}^2}{m_{2k} \omega_{2k}} \delta(\Omega-\omega_{2k}) ~.\label{spectralfunctiondef}
\end{align}
The identity in Eq.~(\ref{Z}) with the Lagrangian in~(\ref{lagrangian}) holds  if these spectral functions fulfil the following integral equations:
\begin{align}
   \mathcal{K}_{\mu}(\omega) \; = \; \int \! \frac{\mathrm{d} \Omega }{\Omega} \frac{J_{\mu}(\Omega)}{\pi} \frac{\omega^2}{\omega^2 + \Omega^2}~.  \label{K}
\end{align}
It can be checked that this is the case if we choose
\begin{align}
\begin{split}
   J_c(\omega) =&  \frac{\omega}{ \pi K_c} \Theta(\omega)\, , \\
   J_s(\omega) =&  \frac{1}{ \pi K_s} \left[ \sqrt{\omega^2-\Delta^2} ~\Theta(\omega-\Delta) + \frac{\pi}{2} \Delta \omega \delta(\omega) \right] \, . \label{spectralfunctions}
\end{split}   
\end{align}
The tunneling probability to lowest order in the (phenomenologically introduced) 
tunneling matrix element $\gamma$ is obtained using Fermi's golden rule for tunneling between neighboring minima of the potential~\cite{Furusaki_Nagaosa_1993}. 
We consider the minima $(q_c,q_s)=(0,-\frac{\pi}{2})$ and $(\pi,\frac{\pi}{2})$. The probability is given by
\begin{align}
\begin{split} 
 \mathcal{P}_{(0,-\frac{\pi}{2}) \to (\pi,\frac{\pi}{2})} \; =& \;  2 \pi \gamma^2 \sum_{i,f} |\left\langle f | i \right\rangle|^2 e^{-\beta E_i} \delta(E_f-E_i-eV) \bigg/ \sum_{i} e^{- \beta E_i} \\
                       \; =& \; \gamma^2 \int_{-\infty}^{\infty} \! \mathrm{d} t \, \braket{e^{-i H_f t} e^{i H_i t}}_i e^{i e V t} ~ , \label{probability1} \end{split}
\end{align}
where $V$ is the applied voltage, $\beta =1/T$ is the inverse temperature and $|i \rangle$ ($|f \rangle$) represent eigenstates of $H_i$ ($H_f$), defined in Eqs.~(\ref{Hi}) and (\ref{Hf}) below, with eigenvalues $E_i$ ($E_f$). 
The thermal average is defined as 
$\braket{X}_i = \text{Tr}(X e^{-\beta H_i}) / \text{Tr}(e^{-\beta H_i})$. The initial and final state Hamiltonians are obtained from $\mathcal{L}$ in Eq.~(\ref{lagrangian}) by setting $(q_c,q_s)=(0,-\frac{\pi}{2})$ and $(\pi,\frac{\pi}{2})$, respectively. After quantizing the oscillator modes we find
\begin{align}
\begin{split} 
   H_i \; =& \; \sum_j \omega_{1j} (a_j^{\dagger} a_j^{\phantom{\dagger}} + \frac{1}{2}) 
                + \sum_k \Big[ \omega_{2k} (b_k^{\dagger} b_k^{\phantom{\dagger}} +\frac{1}{2}) - \frac{\pi g_{2k}}{2 \sqrt{2 m_{2k} \omega_{2k}}} (b_k^{\phantom{\dagger}}+b_k^{\dagger})  + \frac{\pi^2 g_{2k}^2 }{8 m_{2k} \omega_{2k}^2} \Big] \, , \label{Hi}
\end{split}\\
\begin{split} 
   H_f \; =& \; \sum_j \Big[ \omega_{1j} (a_j^{\dagger} a_j^{\phantom{\dagger}} +\frac{1}{2}) + \frac{\pi g_{1j}}{\sqrt{2 m_{1j} \omega_{1j}}} (a_j^{\phantom{\dagger}}+a_j^{\dagger})  + \frac{\pi^2 g_{1j}^2 }{2 m_{1j} \omega_{1j}^2} \Big] \\
        & + \sum_k \Big[ \omega_{2k} (b_k^{\dagger} b_k^{\phantom{\dagger}} +\frac{1}{2}) + \frac{\pi g_{2k}}{2 \sqrt{2 m_{2k} \omega_{2k}}} (b_k^{\phantom{\dagger}}+b_k^{\dagger})  + \frac{\pi^2 g_{2k}^2 }{8 m_{2k} \omega_{2k}^2} \Big] ~.\label{Hf}
\end{split}           
\end{align} 
The two Hamiltonians are related to each other via the translation of the oscillator coordinates, $x_{1j }\to x_{1j} + \pi g_{1j} / \sqrt{2 m_{1j} \omega_{1j}}$ and
$x_{2k} \to x_{2k} + \pi g_{2k} / \sqrt{2 m_{2k} \omega_{2k}}$, which corresponds to the transformation $H_f = U^{\dagger} H_i U^{\phantom{\dagger}}$ with the unitary translation operator
\begin{align}
   U = \exp\left[ \sum_j \frac{\pi g_{1j}}{\sqrt{2 m_{1j} \omega_{1j}^3 }} (a_j^{\dagger} - a_j^{\phantom{\dagger}}) + \sum_k \frac{\pi g_{2k}}{\sqrt{2 m_{2k} \omega_{2k}^3 }} (b_k^{\dagger} - b_k^{\phantom{\dagger}}) \right] ~.\label{translationoperator}
\end{align}
With the help of this relation the transition probability is evaluated as
\begin{align}
   \mathcal{P}_{(0,-\frac{\pi}{2}) \to (\pi,\frac{\pi}{2})} \; =& \; \gamma^2 \int_{-\infty}^{\infty} \! \mathrm{d} t \, \exp\left[ i e V t - \pi \int_{-\infty}^{\infty} \!  \frac{\mathrm{d} \omega }{\omega^2} \, \big( J_c(\omega) + J_s(\omega) \big) \left[ (1-\cos(\omega t)) \coth(\beta \omega/2) \right]  \right]   ~. 
\end{align} 
In the same way the probability of the reverse process, $(q_c,q_s) = (\pi,\frac{\pi}{2}) \to (0,-\frac{\pi}{2})$ is obtained
\begin{align}
\begin{split} 
   \mathcal{P}_{(\pi,\frac{\pi}{2}) \to (0,-\frac{\pi}{2})} \; =& \; \gamma^2 \int_{-\infty}^{\infty} \! \mathrm{d} t \, \braket{e^{-i H_i t} e^{i H_f t} }_f e^{- i e V t} \\
                                     \; =& \; e^{-\beta e V} \mathcal{P}_{ (0,-\frac{\pi}{2}) \to (\pi,\frac{\pi}{2}) } \, . \end{split}
\end{align}
Here the last line represents the detailed-balance condition. The net current across the impurity potential is given by the difference of the tunneling probabilities
\begin{align}
\begin{split} 
   j_c \; =& \; 2 e ( \mathcal{P}_{ (0,-\frac{\pi}{2}) \to (\pi,\frac{\pi}{2}) } - \mathcal{P}_{(\pi,\frac{\pi}{2}) \to (0,-\frac{\pi}{2})} ) \\
       \; =& \; 2 e \gamma^2 (1-e^{-\beta eV}) \int_{-\infty}^{\infty} \! \mathrm{d} t \, \exp\left[ i eV t- \pi\int \! \frac{\mathrm{d} \omega}{\omega^2} [J_c(\omega) + J_s(\omega)]   e^{-|\omega|/\Delta}\left( (1-\cos\omega t) \coth\frac{\beta \omega}{2} \right) \right] ~,\end{split}
\end{align}
where the factor $2$ in front comes from the spin degeneracy and we introduced the ultraviolet cutoff $e^{-|\omega|/\Delta}$. From this result we obtain the conductance at the voltage $V \to 0$ as a function of the temperature:
\begin{align}
\begin{split}
  G(T) \; =& \; 2 e^2 \gamma^2 \beta  \int_{-\infty}^{\infty} \! \mathrm{d} t \, \exp\left[ - \pi\int \! \frac{\mathrm{d} \omega}{\omega^2} [J_c(\omega) + J_s(\omega)]   e^{-|\omega|/\Delta}\left( (1-\cos\omega t) \coth\frac{\beta \omega}{2} \right) \right] \\
       \simeq & \; 2 e^2  \pi^{3/2} \frac{\Gamma\left(\frac{1}{2 K_c} \right)}{ \Gamma\left(\frac{1}{2 K_c} + \frac{1}{2} \right) } \left(\frac{\gamma}{\Delta}\right)^2 \left( \frac{\pi T}{\Delta} \right)^{\frac{1}{K_c} -2} ~,\label{conductance}
\end{split}       
\end{align}
where $\Gamma(x)$ denotes the Euler gamma function.
\end{widetext}
It is instructive to rewrite the conductance in Eq.(\ref{conductance}) as
\begin{align}
   G(T)\; \propto \; e^2 \left(\frac{\gamma}{\Delta}\right)^2 \left( \frac{\pi T}{\Delta} \right)^{\frac{1}{K_c} -2} \; \equiv \; e^2 \frac{\gamma^2(T)}{\Delta^2}~, \label{tunnelingconductance}
\end{align}
with the renormalized tunneling amplitude at energy $\epsilon$
\begin{align}
   \gamma(\epsilon) =  \left(\frac{ \epsilon }{ \Delta} \right)^{\frac{1}{2 K_c}-1} \gamma ~. \label{renormalizedtunneling}
\end{align}
Equation (\ref{renormalizedtunneling}) shows that the tunneling is enhanced in regime II, as long as interactions are not too strong, i.e., as long as $K_c>1/2$. Thus the conductance  increases in region II as temperature is decreased, as shown in Fig.~\ref{Fig:conductance_schematic}. The physical reason for this behavior will be discussed in detail in the following sections. 
For now let us just point out that the spin Luttinger parameter $K_s$ does not appear in the exponent in Eq.~(\ref{renormalizedtunneling}). As has been discussed above, this is because the spin degrees of freedom in the bulk are gapped and thus do not contribute to the physics at energies below the gap. 

We have checked explicitly that the topologically trivial CDW phase shows a different behavior of the conductance. In this phase the conductance to order $\gamma^2$ is exponentially suppressed as $G \propto t^2 \exp(-\Delta/T)$ for temperatures below the spin gap and shows the usual Luttinger liquid behavior $G \propto T^{\frac{1}{K_c}+\frac{1}{K_s}-2}$ for temperatures above the gap. The main contribution to the conductance at $T \ll \Delta$ in this phase is due to tunneling of singlet pairs and arises only in the order  $\mathcal{O}(\gamma^4)$ in the perturbative expansion of the conductance.

To gain a better understanding of the physics governing transport in the regime II (strong barrier) of the topologically gapped phase, we next study the edge state spectrum and the single particle density of states in this regime.


\section{Boundary bound state}
\label{sec:Boundary bound state}


In the preceding Section we have seen that, unlike in a disordered Luttinger liquid, the strong impurity fixed point is unstable in the presence of weak tunneling if the spin sector is  topologically gapped. As we will see, this is crucially related to the boundary states that emerge at the impurity position in the presence of a bulk gap. We refer the reader to Ref.~\cite{Starykh_2000} for a related discussion of the effect of boundary states on the transport in two-subband quantum wires.

In this Section, we discuss the properties of the boundary bound state of the model in Eqs.~(\ref{Hc}) - (\ref{Himp}).
This is most conveniently achieved by mapping the model to two copies of the boundary sine-Gordon model.

In the strong impurity regime, $g_b /v_s\gg 1$, the charge and spin fields at the origin develop expectation values which minimize the potential energy of the impurity, Eq.~(\ref{Himp}), $E_\text{imp} \propto \cos\Phi_s(0)\, \cos\Phi_c(0) $.  This is minimised by any combination $\Phi_c(0)=n\pi$, $\Phi_s(0)=m\pi$ where $n+m$ is an even integer. The discussion which follows is analogous for any of these degenerate minima, so to be concrete, we choose $n=m=0$, i.e. $\Phi_c(0)=0$ and $\Phi_s(0)=0$.

While the above discussion is correct for the charge mode as there is no other term competing with this minima, this is not the case for the spin mode, where the bulk sine-Gordon term in Eq.~\eqref{Hs} which is also in the strong coupling regime is minimised by a different value of $\Phi_s$.  We therefore must study the spin sector more closely.  By integrating out the small fluctuations of the charge field around the mean field and redefining the spin fields as $\Phi_s \to (\Phi_s - \pi)/2$ the model maps to two copies of the boundary sine Gordon model with the action
\begin{equation}
S_s = S_{s,1} + S_{s,2}
\label{action-Ss}
\end{equation}
with 
\begin{align}
\begin{split} 
   S_{s,1} \, =& \, \frac{v_s}{16 \pi K_s}  \int_{-\infty}^0 \! \mathrm{d} x \int \mathrm{d} \tau \,  \Big[ v_s^{-2} ( \partial_{\tau} \Phi_{s } )^2 + (\partial_x \Phi_{s})^2  \Big]\\
   & \,   -    \frac{g_{\text{sG}} }{ (2 \pi a)^2} \int_{-\infty}^0 \! \mathrm{d} x  \int \mathrm{d} \tau \,     \cos  \Phi_{s}  \\
   &  \, - \frac{g_b}{2 \pi a} \int \! \mathrm{d} \tau \, \cos \left. \left( \frac{1}{2} \left( \Phi_{s}   - \pi \right)\right)\right|_{x=0}\, . \label{bsGaction}
\end{split}   
\end{align}
The action $S_{s,2}$ is defined identically to Eq.~(\ref{bsGaction}) but for fields with coordinates $x>0$.
It has been shown~\cite{Schuricht_2011} that the model in (\ref{bsGaction}) supports boundary bound states with energy
\begin{align}
   E_{\text{BBS}} = \Delta \sin \chi, \quad \chi = \frac{\pi-\Phi_s^0}{2-2 K_s} \, , \label{BBSenergy}
\end{align}
where $\Phi_s^0 = \Phi_s(0,\tau)$ denotes the value of the mean field solution at the origin.
In particular, in the case of fixed boundary conditions, which can be obtained from (\ref{bsGaction}) in the limit $g_{\text{b}} \to \infty$, the mean field takes the value $\Phi_s^0 =\Phi_s(x=0,\tau) =  \pi$ and thus the energy of the boundary bound state (\ref{BBSenergy}) is exactly zero. This was also discussed in previous works in a different framework \cite{Kainaris_Carr_2015, Keselman_2015}.

The physical nature of the bound state has been discussed by Ghoshal and Zamolodchikov~\cite{Ghoshal_1994, *GhoshalE_1994}.
For $0< \Phi_s^0 < \pi$ the ground state of each boundary sine-Gordon model is characterized by the asymptotic behavior $\Phi_s^{(1)} \to 0$ as $x \to \pm \infty$. Classically, there exists another stable state with $\Phi_s^{(2)} \to 2 \pi$ as $x \to \pm \infty$. This state is expected to be stable in the quantum theory as well if the parameter $\Phi_s^0$ is not too small (compared with the parameter $\sqrt{K_s}$ governing quantum fluctuations). Exactly for $\Phi_s^0 = \pi$ both ground states are degenerate and the energy of the boundary bound state vanishes. In this scenario the bound state can emit or absorb a soliton changing its state between the two degenerate ground states without energy cost. This flipping between degenerate edge state configurations allows single electrons to tunnel across the barrier although the bulk spin sector has an excitation gap for single spins.

Having understood the $g_b \to \infty$ limit, we turn to a more physical setup with a finite potential barrier due to the impurity, which is equivalent to a small but finite tunneling amplitude $\gamma$ for electrons. 
Let us first study how a finite barrier strength affects the classical ground state of (\ref{bsGaction}). By minimizing the action $S_s$, Eqs.~(\ref{action-Ss}) and (\ref{bsGaction}), we obtain the equations of motion
\begin{align}
   \partial_y^2 \Phi_s = \sin \Phi_s - \bar{g}_b \delta(y) \cos \frac{\Phi_s}{2} \, , \label{eqofmotion}
\end{align}
with dimensionless coordinate $y= (2 K_s g_{\text{sG}} / v_s  \pi a^2)^{\frac{1}{2}} x$ and the dimensionless parameter $\bar{g}_b = (8 \pi v_s K_s/g_{\text{sG}})^{\frac{1}{2}} g_b$.
We solve Eq.~(\ref{eqofmotion}) by an appropriate ansatz. For $y>0$ and $y<0$ the solution should have a form of the bulk soliton. Requiring  the asymptotic condition $\Phi^{(2)}_s(y) \to 2 \pi$ as $|y| \to \infty$, we get
\begin{align}
   \Phi_s^{(2)}(y) = \begin{cases} 4 \arctan e^{-y-y_1} \, &, \enspace y<0~, \\
                           4 \arctan e^{y-y_2} \, &, \enspace y>0    ~.         \end{cases}
\end{align}
The constants $y_1$ and $y_2$ are determined by the matching conditions at $y=0$:
\begin{align}
   \Phi_s^{(2)}(0+) &= \Phi_s^{(2)}(0-) ~,\\
   \partial_y \Phi_s^{(2)}(0+) - \partial_y \Phi_s^{(2)}(0-) &= - \bar{g}_b \cos \frac{\Phi_s^{(2)}(0)}{2} \, .
\end{align}
The first condition is simply the continuity of the solution and the second is obtained by integrating the equation of motion (\ref{eqofmotion}) over an infinitesimal interval around the origin. 
Applying these conditions, we find $y_1 = y_2 = \text{arsinh}(4/\bar{g}_b)$. In particular, in the strong-barrier limit, $\bar{g}_b \gg 1$, we obtain $\Phi_s^{(2)}(x=0) \simeq \pi - 8/ \bar{g}_b$, and thus  the energy of the bound state (\ref{BBSenergy}) takes the form
\begin{align}
   E_{\text{BBS}}^{(2)} = \Delta \sqrt{\frac{2 g_{\text{sG}}}{\pi K_s (1-K_s)^2}} \frac{1}{g_b} ~.
\end{align}
The calculation for the solution with asymptotics $\Phi_s^{(1)} \to 0$ is analogous and yields $E_{\text{BBS}}^{(1)} = E_{\text{BBS}}^{(2)}$. 

From the problem of tunneling across a delta-function potential barrier for noninteracting particles we know that the tunneling amplitude is inversely proportional to the barrier strength. Interactions renormalize the tunneling but do not change this relation. Thus, we conclude that, in the presence of weak tunneling through the barrier and at a finite temperature, the renormalized energy of the boundary bound state scales as
\begin{align}
   E_{\text{BBS}}(\epsilon) \propto |g_b(\epsilon)|^{-1} \propto |\gamma(\epsilon)| \propto \epsilon^{\frac{1}{2 K_c}-1} ~, \label{EBBS}
\end{align}
where the scaling of $\gamma(\epsilon)$ was determined in Eq.~(\ref{renormalizedtunneling}). The linear scaling of $E_{\text{BBS}}$ with 
the renormalized tunneling amplitude $\gamma(\epsilon)$ can be also understood from a simple physical reasoning. For an infinite barrier ($g_b = \infty$, $\gamma = 0$), there are zero-energy bound states on each side of the barrier. At finite (but small) $\gamma(\epsilon)$, they get split acquiring an energy proportional to the renormalized matrix element  $\gamma(\epsilon)$. Hence, if $K_c>1/2$, the energy splitting grows according to Eq.~(\ref{EBBS}) as temperature is decreased. One observable where this splitting can be directly seen experimentally is the local density of states, which will be discussed in the next Section.

\section{Local density of states}
\label{Sec:Local density of states}

As we have discussed in the previous Sections, the transport in the regime II of Fig.~\ref{Fig:conductance_schematic} is governed by bound-state-mediated single-electron tunneling which leads to an increase of conductance as temperature is lowered. On the other hand, we know that at lowest temperatures (regime I), where the conductance is nearly perfect (ballistic), the transport is governed by pair tunneling of spin singlets. In order to study the crossovers between the different transport regimes, we now consider the local density of states in the regimes I-III. 
 
\begin{figure*}
  \centering
   \includegraphics[width=.8\textwidth]{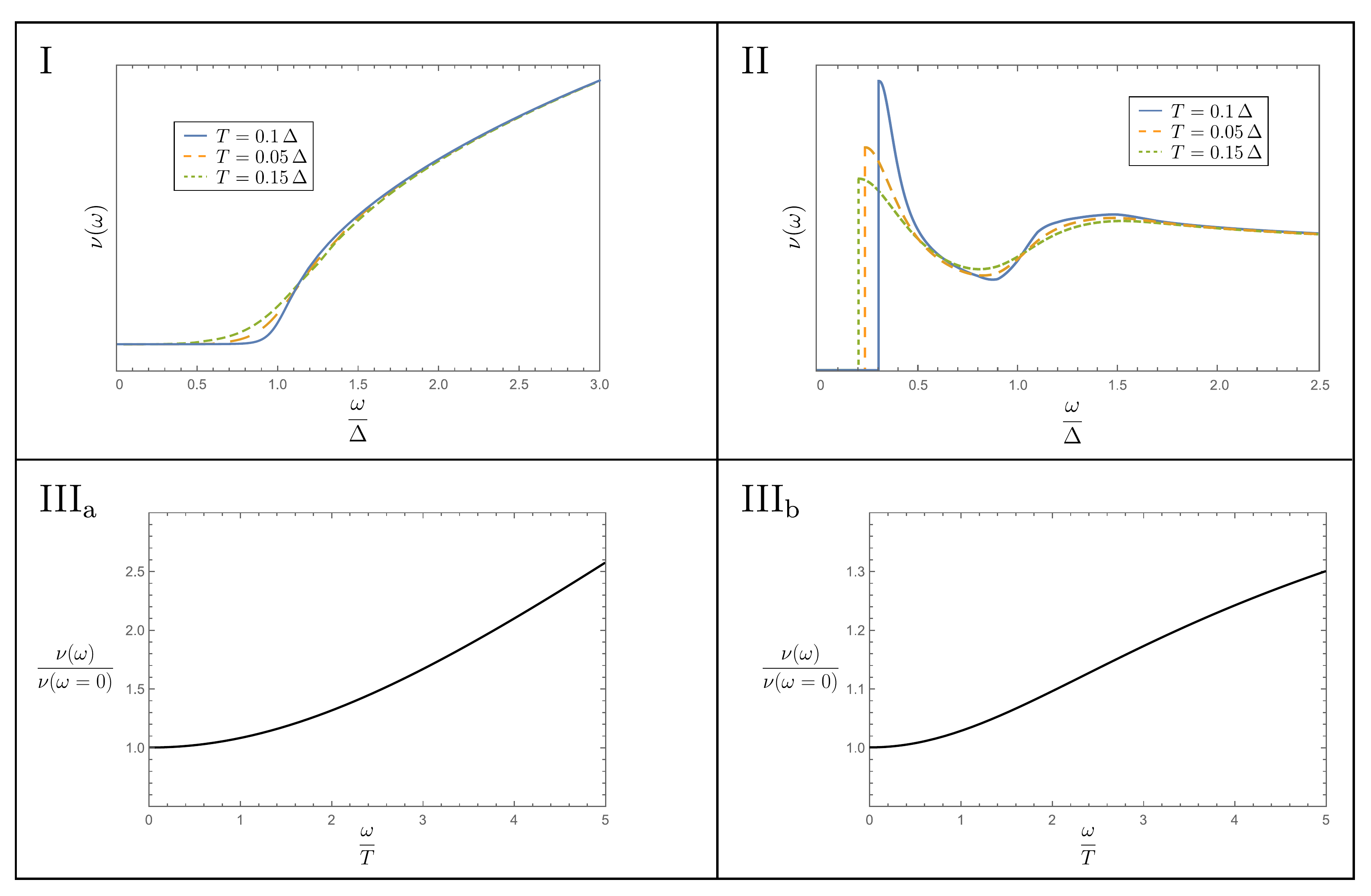}
  \caption{Plot of the LDOS (at the position of the impurity) for a spin-anisotropic 1D electron gas in the presence of a single impurity in the different transport regimes of Fig.~\ref{Fig:conductance_schematic}. Parameters are chosen as $K_c = .8$ and $K_s = .5$. Regime III corresponds to the Luttinger liquid phase. In the regimes $\text{III}_{\text{a}}$ and $\text{III}_{\text{b}}$ the relevant energy scales are much larger than $\Delta$ so that the value of the gap is irrelevant for the discussion of the LDOS. The density of states shows a power-law behavior with a zero bias anomaly that is cut off by the finite temperature. The power law exponent depends on whether the impurity has flown to strong coupling ($\text{III}_\text{b}$) or stays at weak coupling ($\text{III}_\text{a}$). In Regime II the system is in the gapped SDW phase with a strong impurity. The LDOS shows a subgap contribution due to the presence of topological edge states at the impurity position. In regime II we only focus on frequencies $\omega \lesssim \Delta$ (at $\omega \gg \Delta$ contributions that have not been included will produce conventional Luttinger liquid behavior). Regime I shows the LDOS for the SDW phase with a weak impurity.  }
  \label{Fig:DOSoverview}  
\end{figure*}

The observable of main interest in this Section is the local tunneling density of states of electrons at the impurity position, which is defined as
\begin{align}
   \nu(\omega) = - \frac{1}{\pi} \lim_{i \omega_n \to \omega + i 0+}  \text{Im} \!\! \sum_{\sigma=\uparrow,\downarrow} \left. G_{\sigma}(x_1,x_2,\omega_n) \right|_{x_1=x_2 = 0-}.
   \label{DOSdef}
\end{align}
Here $\omega_n$ are fermionic Matsubara frequencies, and the Green's function of electrons is given by
\begin{align}
\begin{split} 
  & G_{\sigma} \; = \; e^{i k_F (x_1-x_2)} G_{\sigma}^{RR} 
                                  + e^{-i k_F (x_1-x_2)}G_{\sigma}^{LL} \\
                                  & \hspace{1.3cm} +  e^{i k_F (x_1+x_2)} G_{\sigma}^{RL} + e^{-i k_F (x_1+x_2)}G_{\sigma}^{LR}, \label{Greenfunction}
\end{split}                                  
\end{align}
with the chiral fermionic Green's functions
\begin{align}
   G_{\sigma}^{\eta_1 \eta_2} = - \bra{0} T_{\tau} \psi_{\eta_1,\sigma}(x_1,\tau)  
   \psi_{\eta_2,\sigma}^{\dagger}(x_2,0)  \ket{0}~, \label{chiralGF}
\end{align}
where $\ket{0}$ is the ground state. Upon bosonization, the Green's function factorizes into a product of correlation functions in the charge and spin sectors, yielding in the $x \to 0$ limit
\begin{align}
    G_{\sigma}(0,0,\tau) =& - \frac{1}{2 \pi a} g_c( \tau) g_s( \tau) ~.\label{electronGF}
\end{align}  
In the following we will discuss the form of the Green's function [i.e., of the functions $g_c$ and $g_s$ in Eq.~(\ref{electronGF})] and the resulting LDOS in the regimes I-III of Fig.~\ref{Fig:conductance_schematic}. A schematic plot of the LDOS in the different transport regimes is given in Fig.~\ref{Fig:DOSoverview}.

\subsection{Region III}
\label{Sec:DOS region III}

We begin with the high-temperature regime (region III). In this regime, the system is in the Luttinger liquid phase and the form of the Green's function is well known~\cite{Giamarchi_book}:
\begin{eqnarray}
   G(x,\tau) &=& \frac{4}{(2 \pi a)^{1-\alpha_c-\alpha_s}} \text{sgn}(\tau) \nonumber \\
&\times&    \left[ \frac{T/2v_c}{\sin \pi T |\tau|} \right]^{\alpha_c} \left[ \frac{T/2v_s}{\sin \pi T |\tau|} \right]^{\alpha_s}. 
\label{GFIII}
\end{eqnarray}
Here, the factor 4 in front results from the degeneracy in spin and chiral sectors, and the exponents $\alpha_c$ and $\alpha_s$ are functions of the Luttinger parameters in the charge and spin sector. Their form differs depending on whether we measure the Green's function in the bulk (regime $\text{III}_\text{a}$) or at the boundary (regime $\text{III}_\text{b}$). To be specific they read as $\alpha_{\mu} = (K_{\mu} +K_{\mu}^{-1})/2  $ in regime $\text{III}_\text{a}$ and $\alpha_{\mu} = \frac{1}{2 K_{\mu}} $
in regime $\text{III}_\text{b}$. The density of states is obtained by plugging the Green's function~(\ref{GFIII}) into the definition of the LDOS, Eq.~(\ref{DOSdef}). The resulting expression reads as
\begin{align}
\begin{split}
 & \nu^{\text{III}}(\omega)        = \frac{4}{ \pi^2 v_c} \cos\left( \frac{\pi}{4 K_c}  \right)\left( \frac{2 \pi a T}{ v_s} \right)^{\alpha_s} \left( \frac{2 \pi a T}{ v_c} \right)^{\alpha_c-1} \\
    &\times    \text{Re} \left\lbrace B\left(-i \frac{\omega}{2 \pi T} + \frac{\alpha_c+\alpha_s}{2}, 1-\alpha_c-\alpha_s \right)\right\rbrace  \Theta\big(\omega\big)~,
   \end{split}
\end{align}
where $B(x,y)$ is the Euler beta function. A plot of the LDOS in the regimes IIIa and IIIb is presented in the two lower panels of Fig.~\ref{Fig:DOSoverview}.

\subsection{Region II}
\label{Sec:DOS region II}
In this regime the bulk of the system is gapped and the impurity potential is large. The large impurity potential effectively acts as a boundary potential so that we have to calculate the LDOS with open boundary conditions.
The charge part in~(\ref{electronGF}) can be calculated in this case using standard open-boundary bosonization methods~\cite{Fabrizio_1995, Mattsson_1997, Schuricht_2011}.
The correlation function in the charge sector is that of a gapless Luttinger liquid at a hard-wall boundary:
\begin{align}
 g_c( \tau) = \left[ \frac{\pi a T/v_c}{\sin \pi T \tau} \right]^{\frac{1}{2 K_c}} ~.
\end{align}
On the other hand, the integrability of the sine-Gordon model on the half line allows for a calculation of the correlation functions in the spin sector using the boundary state formalism introduced by Ghoshal and Zamolodchikov~\cite{Ghoshal_1994, GhoshalE_1994} together with a form factor expansion. This procedure has been performed in Ref.~\cite{Schuricht_2011} where the authors calculate the local chiral Green's functions. The correlation function in the spin sector consists of three parts
\begin{align}
    g_s( \tau)\equiv  g_s^0( \tau) + g_s^1(\tau) +  g_s^b(\tau) ~. \label{defgs}
\end{align}

Let us first discuss the first two terms $g_s^0$ and $g_s^1$ which describe one-particle contributions of the form factor expansion. The first term corresponds to the free propagation of a massive (anti-)soliton and the second term describes a single collision of such a particle with the boundary.  Terms that involve a higher number of particles in the intermediate state as well as higher order corrections due to the boundary lead to subleading corrections: since these processes require the excitation of more gapped particles, they take place at higher energies~\cite{Schuricht_2011}. Since we are only interested in frequencies $\omega \lesssim \Delta$ we can discard those terms. 
Explicitly, the terms $g_s^0$ and $g_s^1$  read:
\begin{align}
  g_s^0(\tau) \equiv  \frac{2 Z_1}{\pi} \left[\cos \left(\frac{\pi}{4}\right)  K_{0} (\tau) + K_{1/2} (\tau) \right] ~,   \label{gs0}
\end{align}
and
\begin{align}
\begin{split} 
g_s^1(\tau)  =& Z_1 \Big[  \int \! \frac{\mathrm{d} \theta }{2 \pi} \, K\left(\theta+i \frac{\pi}{2}\right) e^{-\Delta \left|\tau\right| \cosh \theta} \\
   +& \int \! \frac{\mathrm{d} \theta }{2 \pi} \, K\left(\theta+i \frac{\pi}{2}\right) e^{- \Delta \left|\tau\right| \cosh \theta} \\
   +&  e^{-i \frac{ \pi}{4}}  \int \! \frac{\mathrm{d} \theta }{2 \pi} \, K\left(\theta+i  \frac{\pi}{2}\right) e^{-\Delta \left|\tau\right| \cosh \theta} e^{-\theta/2} \\
   +& e^{i \frac{ \pi}{4}}  \int \! \frac{\mathrm{d} \theta }{2 \pi} \, K\left(\theta+i \frac{\pi}{2}\right) e^{- \Delta \left|\tau\right| \cosh \theta} e^{\theta/2} 
    \Big] ~.
\end{split}    
\end{align}
Here, $K_n(x)$ denotes the modified Bessel function of the second kind, $Z_1$ is a normalization constant which was obtained in Ref.~\cite{Lukyanov_2001}, and $K(\theta)$ is the so called boundary reflection amplitude. In particular, at the exactly solvable Luther-Emery point, $K_s=1/2$, this function is given by $K(\theta) = \; \text{i} \tanh \frac{\theta}{2}$. We stress that the dependence of the Green's function in~(\ref{defgs}) on $K_s$ is only contained in the form of the reflection amplitude $K(\theta)$, the normalization constants $Z_1$, as well as another constant $B$ to be defined below in the discussion of the bound-state contribution $g_s^b$.

\begin{figure}
  \centering
  \includegraphics[width=.45\textwidth]{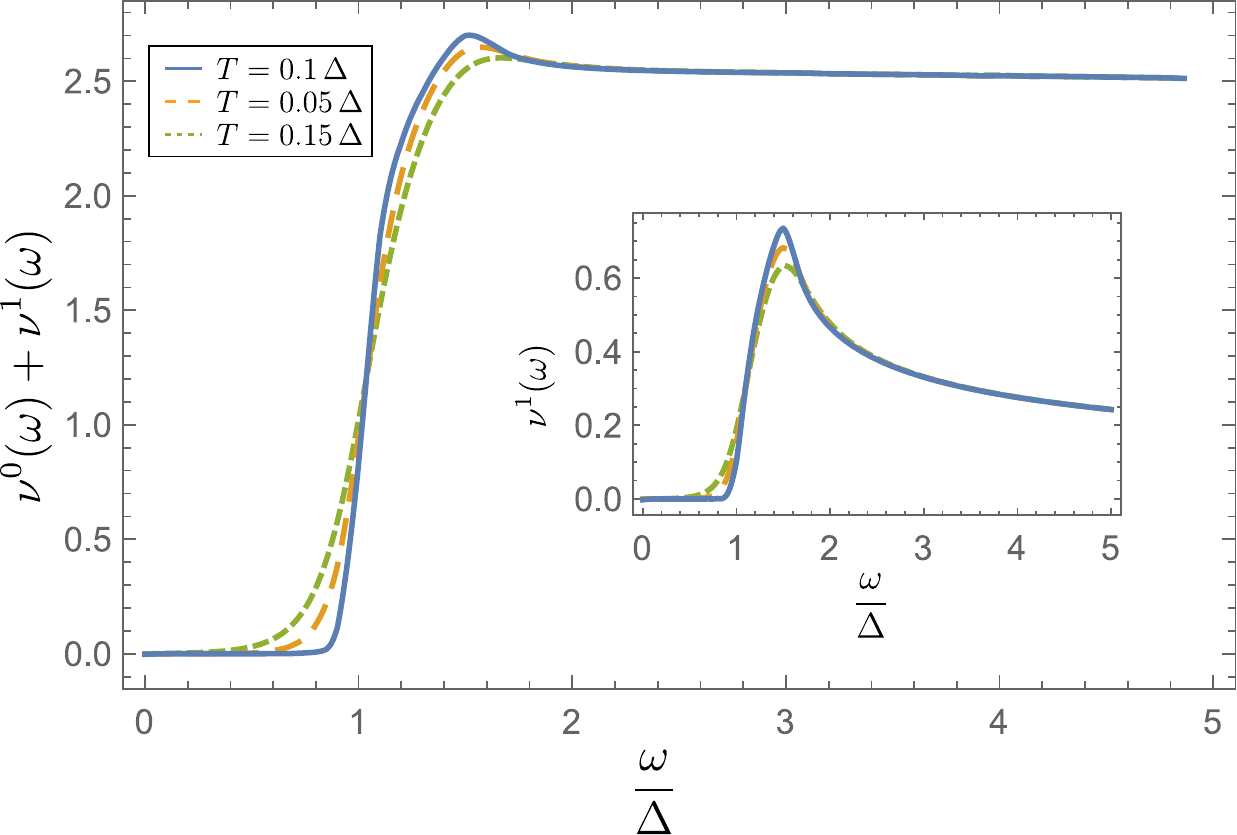}
  \caption{Plot of the LDOS of the bulk states in regime II for $K_c=.8$, $K_s=.5$ for different temperatures. The inset shows the peak structure of the contribution $\nu^1(\omega)$ which arises from the term $g^1$ in Eq.~(\ref{defgs}).} 
  \label{Fig:DOSbulk}
\end{figure}

\begin{figure}
  \centering
  \includegraphics[width=.45\textwidth]{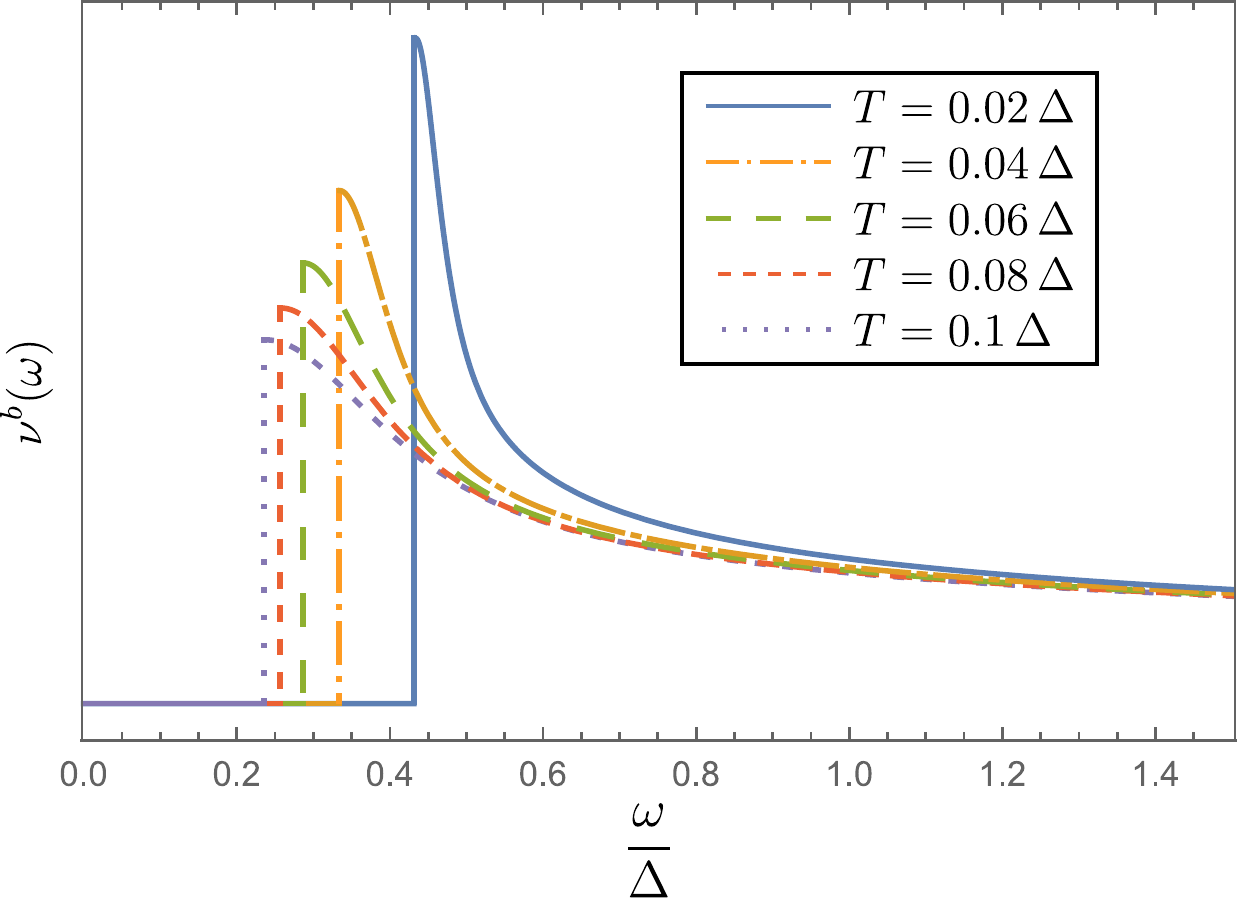}
  \caption{Plot of the LDOS of the edge state in regime II given in Eq.~(\ref{DOSedge}) for different temperatures and for $K_c=.8$. The LDOS vanishes below the edge state energy $E_{\text{BBS}}(T)$ defined in~(\ref{EBBS}). Note that for $K_c>1/2$ this threshold energy increases as temperature is decreased. At the critical temperature when $E_{\text{BBS}}(T^{\ast}) = \Delta$ the contribution disappears into the bulk LDOS signaling the crossover from regime II to regime I. }
  \label{Fig:DOSedge}
\end{figure}

A plot of the bulk contribution to LDOS, which is obtained by only taking into account the terms $g_s^0$ and $g_s^1$ in the Green's function in~(\ref{defgs}),  is shown in Fig.~\ref{Fig:DOSbulk}.
The most salient properties of the bulk LDOS are as follows. First, as a natural manifestation of the gap, the LDOS vanishes exponentially for energies below $\Delta$. Second,  we observe a peak structure at energies just above the gap. We note that the peak is not sharp (i.e., not a $\delta$-function). This is because the electronic Green's function is a convolution of a gapped spin part and a gapless charge part and thus the LDOS is associated with excitations involving at least two ``elementary'' constituents.  Technically, the peak arises due to the contribution $g_s^1$ in~(\ref{defgs}) shown in the inset of Fig.~\ref{Fig:DOSbulk} which describes the propagation of the electron to the boundary where it is reflected and then propagates back to the point of measurement. 

Returning to the full expression for the Green's function, the last term in Eq.~(\ref{defgs}) describes the contribution of the boundary bound state and reads as
\begin{align}
  g_s^b (\tau) \equiv 2 Z_1 B \left[ 1+\cos \frac{\chi}{2} \right] e^{-E_{\text{BBS}} |\tau|}~,
\end{align}
with $E_{\text{BBS}} $ and $\chi$ given in Eq.~(\ref{BBSenergy}), and $B>0$ denoting a real constant. For example at the Luther-Emery point, $K_s = 1/2$, it is given by $B = -2 \cos \Phi_s^0~$.
The contribution to the LDOS at the impurity position arising due to the boundary bound state can be calculated analytically. The calculation is standard~\cite{Giamarchi_book} and yields
\begin{eqnarray}
 \nu^b(\omega)        &=&  \frac{8 Z_1 B}{ \pi^2 v_c} \cos\left( \frac{\pi}{4 K_c}  \right) \Theta\big(\omega-E_{\text{BBS}}\big) \left( \frac{2 \pi a T}{ v_c} \right)^{\frac{1}{2 K_c}-1} \nonumber \\
    &\times&    \text{Re} \left\lbrace B\left(-i \frac{\omega-E_{\text{BBS}}}{2 \pi T} + \frac{1}{4 K_c}, 1-\frac{1}{2 K_c} \right)\right\rbrace .
   \label{DOSedge} 
\end{eqnarray}

The contribution $ \nu^b(\omega) $ arising from the bound state is plotted in Fig.~\ref{Fig:DOSedge} for different temperatures. We observe that the boundary contribution is finite below the bulk gap, vanishing only below the threshold energy $E_{\text{BBS}}$ given by the energy splitting of the edge states. Since the energy splitting increases with lowering temperature according to Eq.~(\ref{renormalizedtunneling}), the threshold for the LDOS shifts towards the bulk gap until it merges with the bulk LDOS at temperatures $\sim T^{\ast}$, where $E_{\text{BBS}}(T^{\ast}) = \Delta$. This temperature scale characterizes the crossover temperature from the edge-state mediated tunneling (at $T^{\ast} < T < \Delta$, regime II) to the singlet pair tunneling (at $T < T^{\ast}$, regime I).
The total LDOS in the regime II, including both the bulk and boundary contributions, is plotted in the upper right panel of Fig.~\ref{Fig:DOSoverview}.

\subsection{Region I}
\label{Sec:DOS region I}

In this regime the bulk of the system is gapped and the impurity potential is weak. Therefore to lowest order we calculate the LDOS in the absence of the impurity potential.

The charge part of the Green's function~(\ref{electronGF}) is thus given by the ``bulk" expression
\begin{align}
 g_c(\tau) = \left[ \frac{\pi a T/v_c}{\sin \pi T \tau} \right]^{\frac{K_c}{2}+ \frac{1}{2K_c}} ~. \label{DOSIc}
\end{align}
and the spin part is given by second term in Eq.~(\ref{gs0}) only:
\begin{align}
  g_s(\tau) \equiv  \frac{2 Z_1}{\pi} K_{1/2} (\Delta \tau)  ~,   \label{DOSIs}
\end{align}
We note that technically this term arises from the RR and LL components of the chiral Green's function in~(\ref{chiralGF}). The offdiagonal chiral components $G^{RL}$ and $G^{LR}$ vanish in regime I, since left and right moving electrons are independent in the absence of a boundary.

The LDOS in this regime can be obtained numerically by using Eqs.~(\ref{electronGF}), (\ref{DOSIc}), and (\ref{DOSIs}).   The results are plotted in the upper left panel of Fig.~\ref{Fig:DOSoverview}.

\section{Conclusion}
\label{Sec:Conclusion}

In this work, we have studied theoretically transport properties of a 1D electron gas with strong correlations  that dynamically gap out the spin degrees of freedom of the low-energy theory, in the presence of a time-reversal invariant impurity. The effective low-energy Hamiltonian of the model is defined by Eqs.~(\ref{Hc})-(\ref{Himp}).
The resulting topological SDW phase is characterized by topologically protected bound states located at the impurity position which carry fractional electron spin.

The key results of this paper concern the behavior of the conductance through the impurity and of the LDOS  in different transport regimes; see Figs. \ref{Fig:conductance_schematic} and \ref{Fig:DOSoverview}, respectively. The results are obtained by using a combination of bosonization and perturbative expansions in different limiting regimes.

At temperatures far above the bulk gap $\Delta$ in the spin sector, i.e., in the regime III of Fig.~\ref{Fig:conductance_schematic}, the conductance  shows behavior typical for a gapless Luttinger liquid in the presence of an impurity. At sufficiently high temperatures, the transport is ballistic with small power-law corrections due to elastic scattering of single electrons off the impurity dressed by Friedel oscillations. The density of states shows a power-law behavior with a zero-bias anomaly that is cut off by the finite temperature. The power-law exponent of both the conductance  and the LDOS depends on whether the impurity has flown to strong coupling or not. A strong impurity effectively corresponds to a boundary and leads to different exponents of the transport observables. The weak-impurity regime is denoted by $\text{III}_\text{a}$ and the strong-impurity regime by $\text{III}_\text{b}$ in the figures.

We have focussed on the range of moderately strong interactions with $1/2  < K_c <1$ and $1/2  \le K_s <1$.
Upon lowering the temperature, both the impurity potential and the soliton interaction potential in the spin sector then flow to strong coupling under the renormalization. The corresponding strong coupling fixed point describes two separate 1D subsystems, each with a gap $\Delta$ for spin excitations. This regime is denoted by II in Fig.~\ref{Fig:conductance_schematic}. The transport in this regime takes place via weak tunneling processes, with amplitude $\gamma$, between the ends of the two subsystems. In view of the topological character of the system, a boundary bound state energetically located within the bulk gap emerges at the end of each subsystem. Due to the finite tunneling between both subsystems, the edge states are energetically split around zero energy by $E_{\text{BBS}}$ defined in Eq.~(\ref{BBSenergy}). 

The dominant transport mechanism in this regime is the single-electron tunneling mediated by the boundary states. Even though single spin excitations are gapped in the bulk, they can be created or annihilated by flipping the edge spin which has an energy cost of order of the splitting. This is clearly visible in the density of states, depicted in Fig.~\ref{Fig:DOSedge}. The DOS has a subgap contribution above a threshold value of $E_{\text{BBS}}$ due to the contribution of the edge state. We note that while the edge state gives a delta function contribution to the DOS in the spin sector, the electron DOS is obtained as a convolution of the DOS of the spin and charge sectors and thus the subgap peak is not sharp.
  
It is important that the energy splitting is not constant but scales with temperature $ \propto |\gamma(T)|$.
Crucially, we find that the tunneling in the regime II is enhanced according to Eq.~(\ref{renormalizedtunneling}). Thus, upon lowering the temperature, the energy splitting of the boundary state gradually increases until finally the edge DOS merges with the bulk DOS at temperature $T^{\ast}$, defined by $E_{\text{BBS}}(T^{\ast}) = \Delta $. Simultaneously, the strength of the impurity potential, which scales  $ \propto |\gamma(T)|^{-1}$, is reduced, ultimately flowing back to a weak-impurity fixed point.  This signals a crossover to a phase where the spin sector is gapped but the impurity potential is weak, denoted by I in Fig.~\ref{Fig:conductance_schematic}. In this regime, the  single electron tunneling is energetically forbidden due to the bulk gap for spin-$1/2$ excitations. This is clearly visible in the DOS in regime I of Fig.~\ref{Fig:DOSoverview}, which shows a hard-gap behavior. The leading transport channel in this regime is then the tunneling of singlet pairs across the impurity. Since this is a much weaker second-order process, the conductance shows ballistic behavior with weak power-law corrections. 

We briefly discuss now what happens if we relax the conditions $1/2  < K_c <1$ and $1/2  \le K_s <1$; see Appendix~\ref{appendix-RG} for a more detailed presentation based on the RG analysis. If $K_c<1/2$, which corresponds to very strong repulsive interactions, pair scattering becomes relevant and the $T=0$ fixed point is insulating.  It is curious that this same limitation also occurs for helical edge states of a two-dimensional topological insulator when interactions are considered \cite{Kainaris_2014}.  If $K_c>1$, which will occur in superconducting realisations of this model \cite{Keselman_2015}, there is no change to regions I and II below the spin gap.  However, if $K_c+K_s>2$, then the impurity is no longer relevant, even above the spin-gap.  This means firstly that the conductance as a function of temperature will be monotonic, and secondly that the regime II where the impurity is still strong at an energy scale of $\Delta$ will be more more difficult to reach.  If $K_s<1/2$, which would correspond to very strong Ising anisotropy, we would expect that the basic physics we have discussed will remain the same, however there may be some quantitative changes due to breather modes in the spin sector that haven't been taken into account in this work. Finally, if $K_s>1$, then generally the system is not in a spin-gapped phase (More accurately, the border is $K_s=1$ only for infinitesimally small backscattering $g_{\text{sG}}$; a finite $g_{\text{sG}}$ slightly shifts the border - see Appendix~\ref{appendix-RG}).

In conclusion, the discussed quasi-long range order SDW phase is an example of a strongly correlated symmetry-protected topological phase that exhibits features fundamentally different from non-interacting topological phases.  We have discussed signatures of these properties both in the LDOS near an impurity and in the behavior of the conductance. We note that although an impurity is irrelevant in the RG sense and will always flow to weak coupling as $T\rightarrow 0$, there are certain parameter regimes where the impurity is strong below the gap, demonstrating boundary states in the LDOS. This physics is rather universal and should be experimentally observable in any of the physical systems listed in Sec.~\ref{Sec:The model}.


A further peculiarity of the system that we have studied is that it has features characteristic for a topological insulator (bulk gap with a topological edge state) only in the spin sector. The charge sector remains gapless. However, as our results show, the charge transport also exhibits remarkable topological properties. Indeed, we have shown that even in the presence of a strong impurity [meaning a (renormalised) impurity strength greater than the gap, implying $G \ll 1$ at intermediate $T$, regime II] the conductance becomes ballistic in the low-temperature limit. Thus, the system combines features of a topological insulator (symmetry-protected topological phase) in the spin sector with those of a topological metal in the charge sector. 

In short, our results on the conductance and the LDOS can be used to experimentally probe the nontrivial topology in the system. We hope that our work will stimulate experimental activity in this direction, both in the condensed-matter and in the cold-atom realizations of the topological phase that we have theoretically explored.

\section{Acknowledgements}
\label{Sec: Acknowledgements}

We thank  M. Bard, E. Berg, I. Gornyi, A. Haim, A. Keselman, and G. M\"oller for useful discussions.  
ADM acknowledges the support within the Weston Visiting Professorship at the Weizmann Institute of Science. This work was supported by the Priority Programme 1666 ``Topological Insulator'' of the Deutsche Forschungsgemeinschaft (DFG-SPP 1666).


\appendix

\section{Renormalization group equations}
\label{appendix-RG}

In this Appendix we develop  the renormalization-group (RG) analysis of the model (\ref{Hc})-(\ref{Himp}) and discuss the corresponding phase diagram in the full range of Luttinger-liquid constants $K_c$ and $K_s$.

The complete action of the model with the Hamiltonian (\ref{Hc})-(\ref{Himp}) reads 
\begin{equation}
S = S_{\text{LL}} + S_{\delta} + S_{\text{SG}} + S_{\text{imp}} +S_{\text{coh}},
\end{equation} 
where
\begin{widetext}
\begin{align}
\begin{split} 
      S_{\text{LL}} \; =& \; \frac{1}{2} \sum_{\mu} \frac{1}{K_{\mu}} \int \! \mathrm{d}^2 r \, \left( \nabla \varphi_{\mu} \right)^2 \, , \\
      S_{\delta} \; =& \; \frac{\delta}{2 K_c} \int \! \mathrm{d}^2 r \,  \left[ \left( \partial_{r_1} \varphi_{c} \right)^2 - \left( \partial_{r_2} \varphi_{c} \right)^2  \right]\, ,
      \\
      S_{\text{SG}} \; =& \; \lambda_{\perp} \int \! \frac{\mathrm{d}^2 r}{a^2} \, \cos(\sqrt{8 \pi} \varphi_s(\bs{r})) \, , \\
      S_{\text{imp}}\; =& \; -\lambda_{\text{imp}}   \int \! \frac{\mathrm{d} r_2}{a} \, \cos(\sqrt{2 \pi} \varphi_s(0,r_2)) \cos(\sqrt{2 \pi} \varphi_c(0,r_2)) \, , \\
      S_{\text{coh}} \; =& \; \int \! \frac{\mathrm{d} r_2}{a} \, \Big[ \lambda_{\text{imp,s}} \cos(\sqrt{8 \pi} \varphi_s(0,r_2)) + \lambda_{\text{imp,c}} \cos(\sqrt{8 \pi} \varphi_c(0,r_2)) \Big] \, . \label{actiondisorder}
\end{split}      
\end{align} 
Here, we defined the dimensionless coupling constants $\lambda_{\perp} = g_{\text{sG}} /( 4 \pi^2 v_s)$ and $\lambda_{\text{imp}} = \, g_b / (\pi v_s)  $, as well as the coordinates $\bs{r} = (r_1,r_2)^T = (x, v_s \tau)^T$. The bosonic fields are related to the convention used in the main text by rescaling $\varphi = \Phi/ \sqrt{2 \pi}$.
The dimensionless velocity difference $\delta = 1- v_c/v_s$, while in principle present, turns out not to be important as it neither flows under RG nor influences any of the other flow equations,
We therefore will simply drop it in the following.  

The action $S_{\text{coh}}$ describes two-particle coherent processes generated by the impurity term in second order in a perturbative expansion in $\lambda_{\text{imp}}$; the bare values of the coupling constants are $\lambda^0_{\text{imp,c}}=\lambda^0_{\text{imp,s}}=0$.
Here, the first term corresponds physically to the backscattering of two incoming electrons with opposite spin, incident from the left and right of the impurity. The resulting scattering process effectively backscatters a particle with spin 1 but zero charge. The second term in $S_{\text{coh}}$ describes a process where two electrons with opposite spin are incident from the same side of the impurity and are coherently backscattered. This process effectively backscatters a singlet with charge $2 e$. These coherent scattering processes become important when either the charge or the spin sector are gapped and electronic excitations are prohibited.

\begin{figure}
	\centering
		\includegraphics[width=0.4\textwidth]{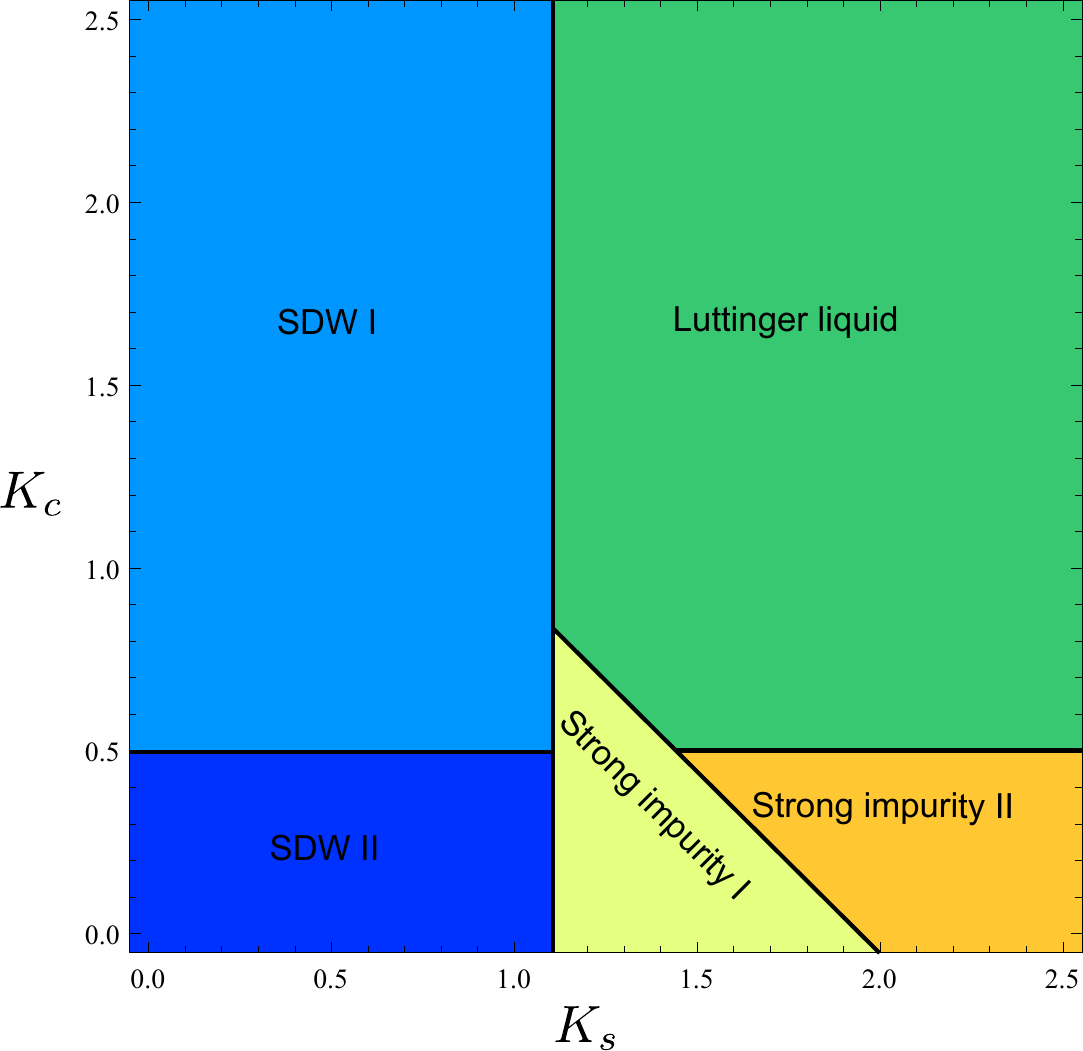}
	\caption{Phase diagram of the model in Eq.~(\ref{actiondisorder}), describing a spin-anisotropic 1D electron gas in the presence of a single impurity. The classification of the phases is defined in Eqs.~(\ref{strongcouplingphases}). The bare parameter of the sine Gordon term is chosen as $\lambda_{\perp}^0=0.2$. The choice of the bare values of the impurity coupling constants does not affect the phase diagram. The phase boundary between the two SDW phases, as well as between the impurity II and Luttinger liquid phase, is $K_c=1/2$. The boundary between the SDW phases and the neighboring phases is $\lambda_{\perp}^0 > 2 (K_s^0-1)$ and the boundary between the impurity I and the impurity II and the Luttinger liquid phase is $K_c^0 + K_s^0 =2$. The present paper focusses on the SDW I phase.
	}\label{Fig:phasediagram}
\end{figure}

The phase diagram of the model in Eq.~(\ref{actiondisorder}) is determined by the interplay of the impurity scattering (described by the terms $S_{\text{imp}}$ and $S_{\text{coh}}$) and the interaction (described by the sine Gordon term). To gain a better understanding of this interplay, we  perform a RG analysis of the action in Eq.~(\ref{actiondisorder}). The RG equations read
\begin{align}
\begin{split} 
   \frac{d K_s}{d \ell} \; =& \; - \frac{1}{2} K_s^2 \lambda_{\perp}^2 \, , \\
   \frac{d \lambda_{\perp}}{d \ell} \; =& \; (2 -2 K_s) \lambda_{\perp} \, , \\
   \frac{d \lambda_{\text{imp}}}{d \ell} \; =& \; \left[ 1 - \frac{1}{2} ( K_s + K_c) \right] \lambda_{\text{imp}} - \frac{1}{2} \lambda_{\text{imp}} \lambda_{\text{imp,c}} - \frac{1}{2} \lambda_{\text{imp}} \lambda_{\text{imp,s}} - \frac{1}{4 \sqrt{2 \pi}} \lambda_{\perp} \lambda_{\text{imp}} \, , \\
   \frac{d \lambda_{\text{imp,s}}}{d \ell} \; =& \; (1 - 2 K_s) \lambda_{\text{imp,s}} - \frac{1}{4} \lambda_{\text{imp}}^2 - \frac{1}{2} \lambda_{\perp} \lambda_{\text{imp,s}}  \, , \\
   \frac{d \lambda_{\text{imp,c}}}{d \ell}\; =& \; (1 - 2 K_c) \lambda_{\text{imp,c}} - \frac{1}{4} \lambda_{\text{imp}}^2 \, . \label{RGequations}\end{split} 
\end{align}
There exists a line of weak-coupling fixed points with $K_s = \text{const}$ and $\lambda_i=0$ for all $i$. The corresponding phase is the spinful Luttinger liquid phase. There is also a number of strong-coupling fixed points:
\begin{align}
\begin{split} 
     \lambda_{\perp} \to 0 \, , \enspace \lambda_{\text{imp,s}} \to 0 \, , \enspace \lambda_{\text{imp,c}} \to 0  \, , \enspace \lambda_{\text{imp}} \to \pm \infty \qquad &\Rightarrow \qquad \text{Strong impurity I} \, , \\  \\
     \lambda_{\perp} \to 0 \, , \enspace \lambda_{\text{imp,s}} \to 0 \, , \enspace \lambda_{\text{imp,c}} \to \infty  \, , \enspace \lambda_{\text{imp}} \to 0 \qquad &\Rightarrow \qquad \text{Strong impurity II} \, ,\\   \\
     \lambda_{\perp} \to 0 \, , \enspace \lambda_{\text{imp,s}} \to \infty \, , \enspace \lambda_{\text{imp,c}} \to 0  \, , \enspace \lambda_{\text{imp}} \to 0 \qquad &\Rightarrow \qquad \text{Strong impurity III}\, , \\ \\
     \lambda_{\perp} \to \infty \, , \enspace \lambda_{\text{imp,s}} \to 0 \, , \enspace \lambda_{\text{imp,c}} \to 0  \, , \enspace \lambda_{\text{imp}} \to 0 \qquad &\Rightarrow \qquad \text{SDW I}\, , \\  \\
     \lambda_{\perp} \to \infty \, , \enspace \lambda_{\text{imp,s}} \to 0 \, , \enspace \lambda_{\text{imp,c}} \to \infty  \, , \enspace \lambda_{\text{imp}} \to 0 \qquad &\Rightarrow \qquad \text{SDW II}\, , \\   \\
     \lambda_{\perp} \to -\infty \, , \enspace \lambda_{\text{imp,s}} \to 0 \, , \enspace \lambda_{\text{imp,c}} \to 0  \, , \enspace \lambda_{\text{imp}} \to \infty \qquad &\Rightarrow \qquad \text{CDW} \, . \label{strongcouplingphases}
\end{split}
\end{align}
Note that the equations for $\lambda_{\perp}$ and $K_s$ decouple from the rest, in the sense that the flow of $\lambda_{\perp}$ and $K_s$ is not influenced by the other couplings. This result is very natural, since the local disorder term cannot affect the physics in the bulk. This can be used to classify the strong-coupling phases above into three strong impurity phases, where the spin gap does not develop and three phases with a gap in the spin sector.

If $\lambda_{\perp}$ flows to zero, the fixed points correspond to those encountered in the study of a single impurity in the Luttinger liquid phase~\cite{Kane_Fisher_1992,Furusaki_Nagaosa_1993}. 
There are then three possible strong-coupling phases. In phase I, the impurity term becomes relevant. Physically, the impurity potential perfectly reflects incoming electrons at zero temperature in the thermodynamic limit and the system is effectively cut into two parts, each being in the Luttinger liquid phase. The impurity phases II and III describe impurity potentials that perfectly transmit spin but no charge, or vice versa (see the discussion in ~\cite{Kane_Fisher_1992,Furusaki_Nagaosa_1993}).
There is, however, one difference between these works and the current discussion. In the presence of the sine Gordon term, the bulk interaction $K_s$ is also subject to renormalization. This renormalization slightly shifts the phase boundaries between the impurity phases in the $K_s$-$K_c$-plane compared to the model with $\lambda_{\perp} = 0$. Since the sine Gordon term is irrelevant in this region of the phase diagram, the shift of the phase boundaries is very minor.

Let us now discuss the opposite situation, when $\lambda_{\perp}$ grows under the RG flow. If the bare parameters of the model obey $|\lambda_{\perp}^0| > 2 (K_s^0-1)$, the flow is towards a strong coupling fixed point where the system dynamically develops a spin gap. The nature of the fixed point then additionally depends on the sign of $\lambda_{\perp}$. For $\lambda_{\perp}<0$, the strong-coupling fixed point is of the CDW type. In this case the development of CDW order in the bulk goes hand in hand with the flow of the impurity to strong coupling. In the thermodynamic limit and at zero temperature, the impurity potential becomes perfectly reflecting and cuts the wire into two parts, each exhibiting a CDW order.

On the other hand, if $\lambda_{\perp}>0$, the strong-coupling fixed point is of the SDW type.
In the SDW phase the impurity potential always renormalizes to zero. Whether the system remains conducting or becomes insulating in the thermodynamic limit then depends on the coupling $\lambda_{\text{imp,c}}$ that is generated by the impurity in second order. We find that the corresponding term becomes relevant for $K_c<1/2$, independently of the physics in the spin sector. Then there are two disordered SDW phases, depending on whether $\lambda_{\text{imp,c}}$ grows or decreases under the flow. In the SDW I phase ($K_c > 1/2$), the system remains a ballistic conductor at zero temperature, while the system in the SDW II phase ($K_c < 1/2$) is insulating.

The overall phase diagram for $\lambda_{\perp}^0>0$ (we have chosen $\lambda^0_{\perp} = 0.2$) is depicted in Fig.~\ref{Fig:phasediagram}. The present paper focusses on the SDW I phase.

\end{widetext}

\bibliographystyle{apsrev4-1}
\bibliography{database}

\end{document}